# Risk Model of German Corona Warning App - Reloaded

Jens Braband, TU Braunschweig; Hendrik Schäbe, TÜV Rheinland

## Abstract

In this paper we discuss the risk model of the German Corona Warning App (CWA) in two versions. Both are based on a general semi-quantitative risk approach that is not state of the art anymore and for some application domains even deprecated.

However, it turns out that the CWA uses a much more limited model, that does not even assess risk, but relies only on one parameter, a weighted exposure time. It is shown that the CWA grossly underestimates even this parameter and so may reassure the users wrongly. As the CWA also has other systematic limitations and shortcomings it is advised not to rely on its results but rather on Covid testing and vaccination.

## Introduction

In Germany, an app has been introduced to cope with the Covid pandemic. The CWA computes a risk for persons to have been infected caused by contacts. In a former paper we analyzed, how CWA works and also its risk model [12]. Recently, after some criticism, the model has been changed (from version 1.9 and later). We study in this paper the changes of the risk model and its relevance as an indicator for individual risk. While the particular discussion focusses on the German app only, the so-called Exposure Notification Framework (ENF), on which the CWA relies, has been introduced by Apple and Google as part of a standard interface, and so it can be expected that the results may be generalized [14].

We introduce how the CWA works, then briefly summarize the former risk model, describe the changes. Finally, we compare the risk models and assess the effects of the changes.

## How the app works

First a simplified overview is given in order to understand the risk model. More details are given in [1][2]. A much broader view on contract tracing apps has been published recently [14].

After a user has installed the CWA, each day a new anonymous ID is created. About every five minutes the environment is scanned for Bluetooth signals emitted from other mobiles. Data like ID, signal attenuation, duration etc. are collected and aggregated for each day.

If the user receives a positive test result and agrees to publish it, then the associated anonymous IDs for the preceding 14 days are transmitted to the central server, from which it is transmitted to all subscribers and checked against the recorded data in the local app. The actual risk evaluation is performed decentralized by each CWA.

# The Basic Risk Model

The basic model up to version 1.7.1 is defined by four parameters [2], which in a first step are evaluated on a semi-quantitative scale each ranging from 0-8 for each day for each ID that reported a positive test result (see figure 1):

- The Days since Exposure (DE) is the time since exposure to the infected person, a value between 0 and 14, durations longer than 14 days are not taken into account.
- The Exposure Duration (ED) is the cumulative time of exposure on the day, takes values between 0 and 8
- The Bluetooth Signal Attenuation (SA) is used as a measure of the distance to the infected person, takes values between 0 and 8
- The Transmission Risk (TR) estimates the level of infectiosity of the person on that day, takes values between 0 and 8

Then the Total Risk Score (TRS) is evaluated by multiplication of DE, ED, SA and TR, theoretically resulting in scores between 0 and 7168. ENF is a standard interface for CWA introduced by Apple and Google, but the parameters can be chosen by each implementation, and so they differ country by country and app version by app version.

ENF resembles the approach known as Risk Priority Numbers (RPN) and suffers from the same limitations and flaws, which are known for about two decades [5][6]. For some application sectors the use of RPN is even deprecated [7]. Generally, such an approach is not considered state of the art anymore and should be used with great care.

The major problem is that the scores for the parameters are often only ordinal scale or rank numbers, for which operations like multiplication or division are not well-defined. As a consequence, the results may lead to under- or overestimation of the related risk [8].

However, in the practical implementation of the CWA the model is simplified, and the coinciding ranges are limited [3] by

- Days since Exposure (DE) is set to 5 for values below 14 days, and 0 for above, leading to $\delta_{DE} = 5*I(DE≤14)$
- Exposure Duration (ED) is set to 0 for all values up to 10 minutes, and 1 for above, yielding $\delta_{ED} = (ED>10)$
- Signal Attenuation (SA) is set to 0 above 73 dB, and 2 for all values below, i. e. $\delta_{SA} = 2* I(SA≤73dB)$
- Transmission Risk (TR) is set to (6, 8, 8, 8, 5, 3, 1, 1, 1, 1, 1, 1, 1) [4], depending on DE, e. g. TR is 6 if DE=1, 8 if DE=2 etc, and 0, if DE=14 or above

Here, I(.) denotes the indicator function, which take value 1 if the expression in brackets holds true, and zero otherwise.

So, most of the parameters are only used as binary indicator variables and in the current configuration the Total Risk Score for a particular day and a particular ID is given by

$$TRS = 10 \cdot \delta_{ED} \cdot \delta_{SA} \cdot \delta_{DE} TR \quad (1)$$

So, with the implementation [3] until version 1.7.1 there are only six possible scores: 0, 10, 30, 50, 60, 80. But also a Minimum Risk Score (MRS) of 11 is defined and all risks below are discarded. But the parametrization of the CWA may be changed.

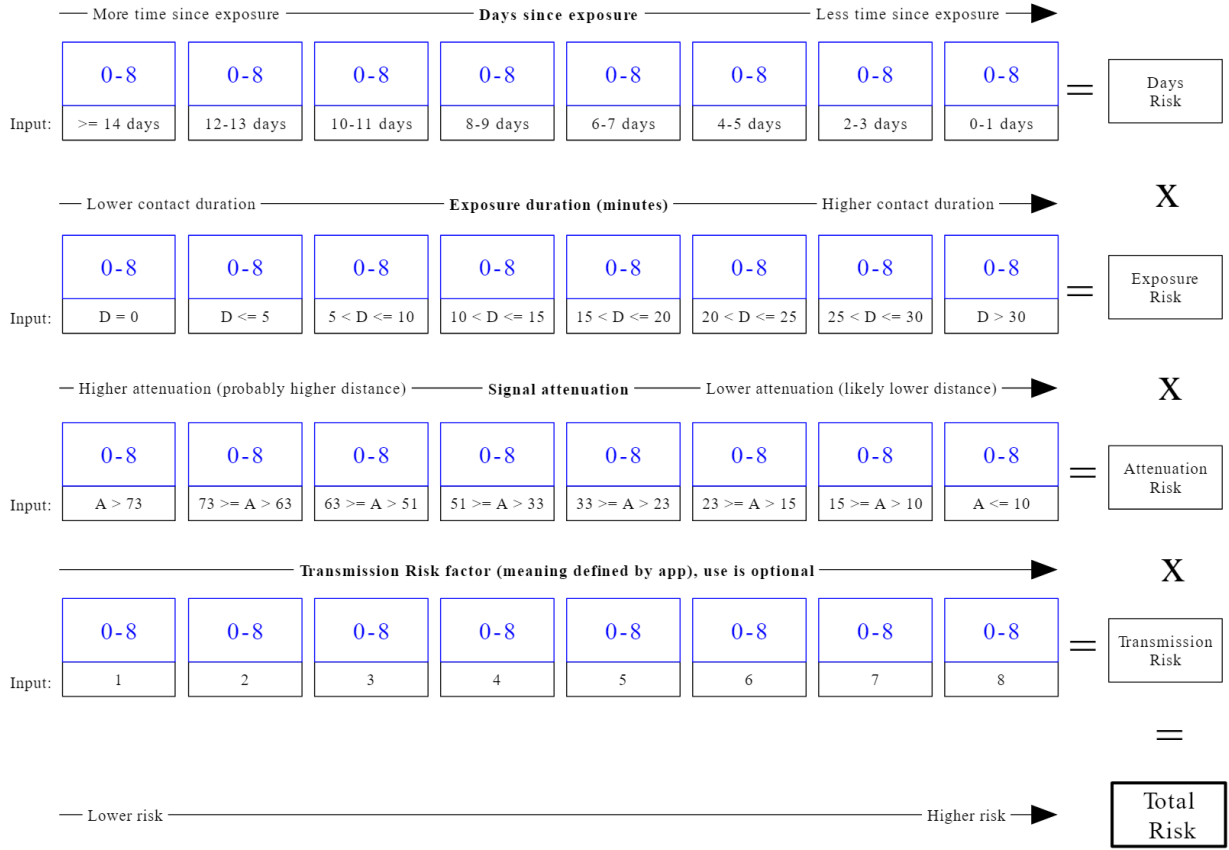

Figure 12: Risk Calculation

Figure 1: Basic Risk Calculation [2]

So, we can conclude that the basic risk model as implemented until version 1.7.1 is more a dosimetric model, depending on the estimated virus concentration, rather than on exposure and other parameters (but for some threshold values). It is not even a risk model as per the definition of many standards. Moreover, the risk model has been heavily discretized.

Example:

- Alice receives a positive test result on the 20th of the month, which she reports immediately.
- Bob is often taking the same bus as Alice. A ride takes 10 minutes, and he has met her on the 16th (two rides) and the 9th (one ride). They have set together with a distance of about 1m.
- For the 16th DE=4 and so TR=8. Both SA and ED are above the threshold and set to 2 and 1, respectively, so TRS=80.

- For the 9th DE=11 and so TR=1. However, ED is below the threshold and set to 0. So TRS=0. Otherwise, the TRS would have been 5, which is below the MRS and would have been discarded anyhow.

## Combined Risk Model

In a second step each CWA combines the scores for different encounters calculated by the basic risk model. Let $R_1, R_2, \ldots R_n$ denote the individual TRS for different days and different IDs that are above the MRS.

In a first step the maximum value $R_{max}$ of the different TRS is determined. Then the ED of all the n encounters is summed up in three different classes: close, medium and far. Let these durations be $t_1$, $t_2$ and $t_3$. For each of the classes a weight is defined and additionally a weight offset, which are denoted by $w_1$, $w_2$, $w_3$ and $w_4$, respectively. Note that in practice the weights for the close and medium classes outweigh the others. Also, an Average Risk Score (ARS), currently 50 [3], is defined. Then the Total Combined Risk (TCR) is calculated as (see figure 2)

$$TCR = (t_1 w_1 + t_2 w_2 + t_3 w_3 + w_4) \frac{R_{max}}{ARS} \quad (2)$$

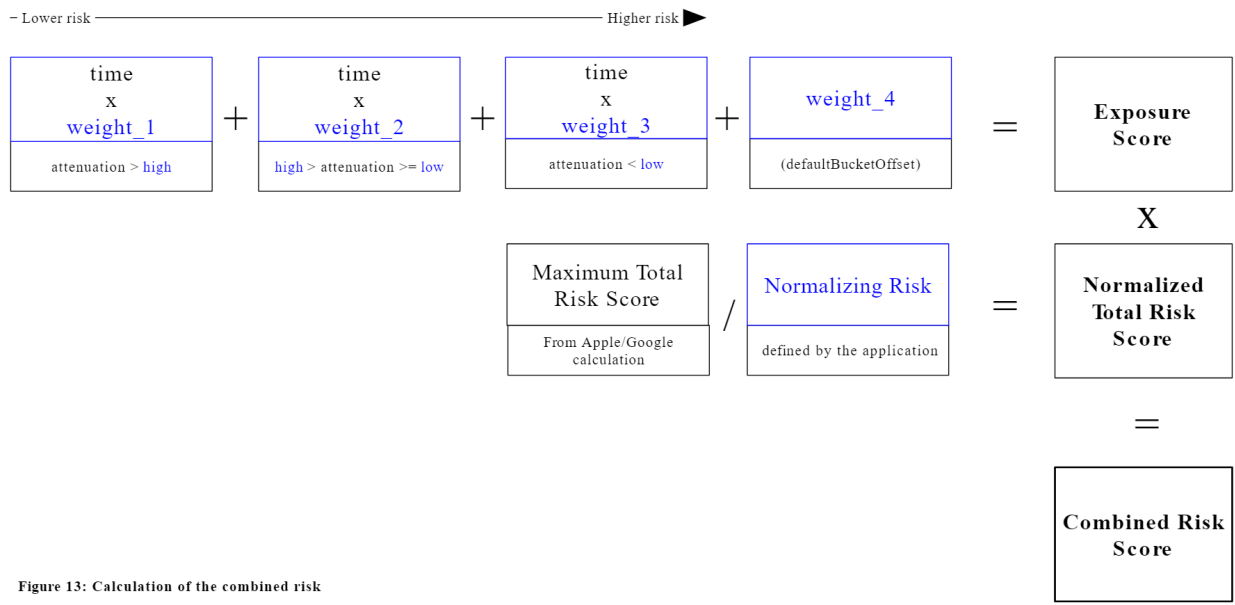

Figure 13: Calculation of the combined risk

Figure 2: Combined Risk Calculation [2]

Surprisingly the TCR is in fact not a risk, but an exposure as the result is in minutes. The first term (in brackets) is a weighted exposure time which is adjusted by a relative factor (dimensionless) which depends on the maximum virus concentration compared with some average.

So overall, without full mathematical exactness, we can characterize the approach used until version 1.7.1 [3] by the German Corona warning app as

$$TCR = (t_1w_1 + t_2w_2 + t_3w_3 + w_4)\frac{max\{10 \cdot \delta_{ED} \cdot \delta_{DE}\delta_{SA} \cdot TR\}}{ARS} \approx (t_1 + t_2/2)\frac{10max\{TR\}}{ARS} \quad (3)$$

Basically, this is not a risk in the narrow sense, it is a weighted exposure duration, the units of TCR are not expected damage per time or similar, but just minutes of exposure. And the weight applied is just a measure of relative infectiosity, which expresses the size of TR relative to a normalizing factor of 5, which is assumed as average infectiosity. And the impact of the factor is limited, the highest possible value being 1.6.

And, if we take additionally into account the uncertainty and spread of all the input parameters, e. g. noise in the signal attenuation, uncertainty in the exposure duration and infectiosity [4], or arbitrariness of the weights chosen, then the model boils down to a quite simple formula and decision procedure:

$$TCR \approx \frac{maxTR}{5}\sum ED \quad (4)$$

- Estimate the minutes that the person was exposed closely to infected persons ($\sum$ED)
- Weight the exposure ED by the infectiosity of the most infected person (max TR/5)
- Take action if the result is HIGH

Example (continued)

- Additionally, Charlie has received a positive test result on the 20$^{th}$, but he reported it only on the 21$^{st}$. But he has installed the app only a week ago.
- He has been on the same bus on the same days, but with some larger distance to Bob (2m)
- For the 16$^{th}$, DE=5 (evaluated on the 21$^{st}$) and so TR=5. Both SA and ED are above the threshold and set to 2 and 1, so TRS=50.
- For the 9$^{th}$, Charlie had not installed the app yet, so there are no data.
- As Alice sat close to Bob, and Charlie in medium distance, $t_1$=$t_2$=20 minutes. The weights are currently set [3] to $w_1$=1, $w_2$=0.5, $w_3$=$w_4$=0, and so the weighted sum results in 30 minutes
- So the TCR amounts to 30 x 80 / 50, which gives 48 minutes.
- The warnings of the app are issued based on the TCR, which are configured [3] as LOW for values up to 15, and HIGH from 15 onwards. So finally Bob would get a HIGH risk warning.

Note that the TCR is almost independent of the distance measured by Bluetooth Signal Attenuation. There is only a loose threshold defined and the exposure duration is weighted into two distance classes. However, it is known, see e. g. the FAQ by RKI [9], that transmssion is mainly by aerosols, and the risk increases when the exposure distance decreases. Close contact to infected persons is also known to be a factor in superspreading events. The importance of the distance is also supported by the fact that it is a major factor in the German Corona rules AHA, where the first letter stands for Abstand=distance.

# Updated Risk Model

From version 1.9 on the risk model has been changed several times [10], with the main intent of improving the granularity of the risk assessment and also the accumulation of risk i. e. several low-risk encounters in the former model may constitute a high risk in the new model. As a caveat it should be noted that some parameters seem to have been adapted several times and that even on the GitHub repository [11] the parameters in the documentation are not consistent. The information presented here is from end of March 2021 and in doubt the authors have used the parameters directly from the configuration files of the repository rather than from the documentation.

As a main means to achieve this 30-minute evaluation windows are introduced, and a contact is only counted if within this window

1. The signal attenuation (SA) is below the threshold of 73 dB for at least 5 minutes (ED), and
2. Transmission Risk (TR) level must be at least 3 (compare figure 1)

If these criteria are not met, then the contact is discarded. Compared to formula (1) this seems not a big change but for the second condition, which was no present before.

Close encounters with SA below 55dB will be counted full, while all other encounters will be discounted by 50%. Note that this does not match with the interface defined by figure 1, but the weight used are the same as before, compare formula (2).

For the TR level a new weighting is introduced, ranging from 0.6 to 1.6, see figure 3. This seems to be a shortcut compared to the calculations in formula (2). Most of the factors are the same as before, e. g. for TR=8, 6, 5, 3, but some have been discarded or slightly changed.

Figure 3: Direct mapping of TRL to scaling factor [11]

Finally, the measured encounter times (more that 5 minutes within a 30-minute window) are weighted by their proximity and multiplied by the TR level weight (from figure 3) and summed up for a particular day. If the sum is below 15 minutes, but above 5 minutes (some sources suggest 13 minutes [11]) the overall risk is LOW or GREEN, otherwise it is HIGH or RED. This is evaluated per day and the app displays the number of days with low or high risk instead of the numbers of encounters with low or high risk (as up to version 1.7.1).

Taking this all into account, it seems as if the general semi-quantitative framework ENF as defined in figure 1 is not really applied anymore. Instead for each encounter i of a particular day a Transmission Risk Score similar to (1) is evaluated (with some changes to the indicator functions):

$$TRSi = \delta_{ED} \cdot \delta_{SA} \cdot ED \cdot TR, \quad (5)$$

where ED is already understood as the weighted encounter times similar to (2)

$$ED = t_1 w_1 + t_2 w_2 + t_3 w_3 + w_4 \quad (6)$$

Finally, the risk is combined per day by simple summation

$$TCR = \sum_{per\ day} TRS_i \quad (7)$$

Example (reloaded):

- For the encounter with Alice on the 16$^{th}$ DE=4 and so originally TR=8, which by figure 3 is then recoded to TR=1.6. Both SA and ED are above the threshold and the weight is 1 and both times are counted full, respectively, so both encounters score TRS=16 minutes, individually. Summing them up a RED warning of HIGH risk is displayed to Bob for this day.
- For the 9$^{th}$ DE=11 and so TR=1, which is recoded to TR=0. So for this risk TRS=0 even as the other thresholds are fulfilled. Otherwise with the lowest TR=0.6 the TRS would have been 6, which would have been at least a GREEN warning with LOW risk for this day.
- Note that the RED warning is independent from the encounter with Charlie on the 16$^{th}$. For this day Charlies' parameters are the DE=5 with TR=5 recoded to TR=1. As the distance was larger the ED is halved to 5, so both encounters would score TRS=5, summing up to 10 for the day, which standalone would lead to a GREEN warning and combined with Alice's scores would lead to combined score of 26.

If we take into account that some scaling factors have been left out in comparison with the previous approach, e. g. for DE and SA, then the results in the example are quite comparable. The most obvious differences are that

- the term related to max{TR} in formulae (3) and (4) has been simplified and has been substituted by a fixed factor based on TR as in figure 3, and
- that the risks are accumulated and reported per day and not combined for all encounters as in formula (4)

# Discussion

The changes in the risk model are not fundamental, so the general criticism from [12] remains valid. In particular the CWA does not estimate individual risk, but individual exposure, as already the units of the resulting TCR shows, which is [min] and not expected harm or expected severity or similar expressions of risk.

Also, the CWA grossly underestimates the exposure duration for the following reasons:

- There have been about 25 million downloads of CWA so far [13], but it is unknown how many installations are active. If we assume optimistically that 30% of the population would have an active app, then less than 10% of the encounters could be registered, as both parties need to have the app activated. Studies indicate [14] that from 60% coverage of the population by CWA, it could become an effective tool.
- Although it is reported that about 60% of the app users have shared positive test results [13], the absolute number of about 250,000 shared results is again less than 10% of all positive test results (about 2,820,000)
- A large number of short contacts (below 5 minutes), that could be close and infectious, are not detected and registered by the app due to energy saving reasons the app scans only about every 5 minutes its environment

This holds also from an individual user's perspective: a user with active CWA meets only other users with active CWA in less that 30% of the encounters, many short encounters, e. g. in the supermarket or at work, will be discarded and only in 60% of the possible cases other infected users will report their positive results. So, also from the individual perspective the CWA is effective only in about 10% of the cases, or even less.

# The Good, the Bad and the Ugly, or: Summary

The risk model of the German CWA has several interesting, somewhat puzzling and also disturbing properties:

1. It has increased the number of users and is also compatible with apps from some neighboring countries, but its coverage is still rather limited.
2. The main advantage is that the CWA records contacts automatically that would otherwise perhaps not have been noticed by the user. But due to data privacy the user can't learn from the warning as it is not reported when and how long the exposure has been. Therefore, the CWA can also not support contact tracing.
3. Also, the number of positive tests reported is only a small percentage of all positive tests
4. In the narrow definition of many standards, it is not a complete risk model, as it estimates only single parameters of risk, but not a comprehensive risk, more a kind of exposure measure. A partial explanation can be based on the decentralized architecture of the app and the incomplete and inaccurate information it uses
5. Even worse, it does not cover many situations with short exposure, that might still go along with a high dose of virus transmission. This may be due to the pressure by the mobile industry to save energy.

6. Also, it is not an a priori risk assessment in order to decide whether a particular action is acceptable, it assesses risk only a posteriori, when it can't be reduce anymore.
7. And finally, most important, it underestimates the true exposure time or virus dose by a large factor, possibly 5 to 10, so it reassures its users to feel safe if it states no or only a low risk, but in reality, the exposure maybe much higher.

In summary, the authors would recommend not to rely on the results of the CWA due to the many shortcomings of its approach and its risk model, some of which are systematic and can not be improved by further updates. A test or even better, vaccination, should always be preferred. Also, it seems that in the near future, when sufficient Corona tests are available, the CWA will even become obsolete. It is rather recommended to develop an app that may estimate risks a priori as a kind of decision support for its users based on their individual risk profile.